\documentclass[12pt]{JHEP3}
\pdfoutput=1

\usepackage{amssymb}
\usepackage{graphicx}

\makeatletter

\title{SCoPE: An efficient method of Cosmological Parameter Estimation}

\author{Santanu Das \& Tarun Souradeep\\
Inter-University Centre for Astronomy and Astrophysics, Post
Bag 4, Ganeshkhind, Pune 411007, India \\
E-mail: \email{santanud@iucaa.ernet.in}, \email{tarun@iucaa.ernet.in}}

\abstract{
Markov Chain Monte Carlo (MCMC) sampler is widely used for cosmological
parameter estimation from CMB and other data. However, due to the
intrinsic serial nature of the MCMC sampler, convergence is often
very slow. Here we present a fast and independently written 
Monte Carlo method for cosmological parameter estimation 
named as Slick Cosmological Parameter Estimator (SCoPE), that employs
delayed rejection to increase the acceptance rate of a chain, and
pre-fetching that helps an individual chain to run on parallel CPUs. An 
inter-chain covariance update is also incorporated to prevent clustering of
the chains allowing faster and better mixing of the chains. We
use an adaptive method for covariance calculation to calculate
and update the covariance automatically as the chains progress. 
Our analysis shows that the acceptance 
probability of each step in SCoPE is more than $95\%$ and the 
convergence of the chains are faster. Using SCoPE,
we carry out some cosmological parameter estimations with different
cosmological models using WMAP-9 and Planck results. One of the current
research interests in cosmology is quantifying the nature of dark energy.
We analyze the cosmological parameters from two illustrative commonly 
used parameterisations of dark energy
models. We also asses  primordial helium fraction in the universe can be constrained
by the present CMB data from WMAP-9 and Planck.  
The results from our MCMC analysis on the one hand helps us to understand the workability
of the SCoPE better, on the other hand it provides a completely independent
estimation of cosmological parameters from WMAP-9 and Planck data.
}
\maketitle

\begin{document}
\section{Introduction}
Precision measurements in the cosmological experiments have improved dramatically
in the past few decades. Several ground based and space based high
precision cosmological experiments have been undertaken and many other future
experiments have been proposed. After WMAP-9 and Planck data release, an
ample amount of data is now available in the hands of cosmologists.
The goal of cosmologists is to extract the maximum amount of information
from these data in about the different cosmological parameters. Thus,
%development of 
techniques for robust and efficient estimation of cosmological 
parameters is one of the most important tools needed in the cosmologist's
arsenal. 

Markov Chain Monte-Carlo (MCMC) methods are widely used to sample the multi-dimensional
space of the parameters and estimate the best-fit parameters from the cosmological
dataset. One of the most widely used MCMC algorithm for sampling
the posterior is Metropolis-Hastings (MH) sampler \cite{Hestings1970,Lewis2002,Metropolis1953,Lewis2013}.
However, MH samplers typically require several thousands of model
evolutions and only a fraction of them get accepted. Hence, it is challenging
to apply the algorithm to problems where the model evaluation is computationally
time consuming. Also due to the intrinsic serial nature of the MH
chains, it often takes long time to map the posterior. Therefore,
even if the multi-processor parallel compute clusters are available they are not
utilized efficiently. In this paper, we present an efficient implementation of the MCMC
algorithm, dubbed, SCoPE (Slick Cosmological Parameter Estimator), where an individual chain can also be run in parallel on multiple processors. 
%Thus all the processors in the cluster can be utilized for running the Monte Carlo chains. 

Another major drawback of the MH method is the choice of step-size.
If the step-size in not chosen properly then the rejection rate increases
and the progress of the individual chain becomes slower. The step
size of the MCMC method is chosen using trial and error method. However,
for the cases where the model evolutions are computationally time consuming,
such as in cosmology, this type of trial and error method is computationaly uneconomical.
Therefore, several authors have proposed different statistical methods
for choosing the optimum step-size. An adaptive proposal Monte Carlo
method is proposed by Haario et al.\cite{Haario1999} that uses the history of
the chains to predict the next movement of the chains to improve the
acceptance of the steps. The concept of inter-chain adaptation has
been proposed in \cite{Craiu2009}. Several other theoretical proposals %calculations
for choosing the optimal step size are also available in literature
\cite{Dunkley2005}. 

There are several codes available for cosmological
parameter estimation. Publicly available CosmoMC \cite{Lewis2002,Lewis2013},
AnalizeThis \cite{Doran2004} codes are MCMC code, widely used for
posterior mapping of the cosmological parameters. There are other
codes such as CosmoPSO \cite{Prasad2012}, CosmoHammer \cite{Akeret2012}
which can find the optimum cosmological parameters very fast, however they
failed to sample the posterior fairly. Hence, the statistical quantities 
(mean, variance and covariance etc.) derived 
from the sample cannot readily yield
%using these codes are not 
unbiased estimates of the population mean, variance etc. 
Also CosmoMC uses the local MH algorithm \cite{Doran2004}, %which in long run 
fairly samples the posterior only asymptotically, i.e. practically 
for 'sufficiently' long run. Hence, if the samples runs are not long enough 
%But if lesser sample-points are used then 
the posteriors may not get sampled fairly.
In this work we devise and implement methodological 
modifications to the MCMC technique that lead to better
acceptance rate. The algorithm proposed in this paper is a standard
global MCMC algorithm combined with 
\begin{itemize}
\item A delayed rejection method that allows
us to increase the acceptance rate. 
\item Pre-fetching is incorporated to
make the individual chains faster by computing the likelihood ahead
of time. 
\item An adaptive inter-chain covariance update is also added to
allow the step-sizes to automatically adapt to the optimum value. 
\end{itemize}

As a demonstration, we use SCoPE to carry out parameter estimation in 
different cosmological models including
the `standard' 6-parameter $\Lambda$CDM model. 
There are many reasons to explore well beyond the simple 6-parameter $\Lambda$CDM
model. Comprehensive comparison
calls for an ability to undertake efficient estimation of cosmological parameters, both owing to increase 
parameters or the increased computational expense for 
each sample evolution.
For example recent data from WMAP-9
and Planck comfirm that the power at the low multipoles of the
CMB angular power spectrum is lower than that predicted in the best-fit
standard $\Lambda$CDM model and unlikely to cause by some observational 
artefact. This has motivated the study of broader class of 
inflationary models that have infra-red cutoff or lead to related desired 
features in the primordial power spectrum \cite{Sinha2006,Jain2009}. 
Another interesting cause for this power
deficiency at the low multipoles can be the ISW effect in a modified
the expansion history of the universe \cite{Das2013a}. Then,
it is important to check if any scenerio in the vast space of dark energy models provides
a better fit to the observational data\cite{Das2013}. In this paper, we analyze
as illustration, two standard dark energy models $\--$ The first one is the constant equation
of state dark energy model \cite{Lewis2002,Hannestad2005,Bean2004}
with constant sound speed. The second one is the CPL dark energy
parametrization proposed in \cite{Chevallier2001,Linder2003} with
a linearly varying equation of state. Our analysis shows that 
both the dark energy models provide marginally better fits to the
data than the standard $\Lambda$CDM model. 

Another important subject in cosmology is the primordial Helium abundance, denoted by $Y_{He}$.
A number of researchers have attempted to pin down the Helium fraction using different 
data sets. Though the primordial Helium abundance %though
does not directly affect the perturbations spectrum, it affects the recombination
and re-ionization processes and consequently changes the CMB power spectrum. The theoretical
prediction of primordial Helium abundance from the standard Big Bang
nucleosysthesis (BBN) is $Y_{He}\thickapprox0.24$ \cite{Ade2013,Trotta2004}.
We have carried out the parameter estimation for $Y_{He}$ 
together with other standard $\Lambda$CDM cosmological parameters to asses the constraint
from current CMB data and check if the allowed range is consistent with 
%to check if it can be fixed by present data and if it is
%then does it resemble
the BBN prediction.%value. 
Our analysis shows that
the data from WMAP-9 and Planck can put a fairly tight constraint on the cosmological
Helium fraction, which matches with the theoretical BBN value.

%The code 
SCoPE is a C code written completely independently from scratch. 
The mixing of multiple chains in SCoPE is better and convergence is achieved faster.
The paper is organized as follows. The second section provides a
brief overview of the standard Metropolis Hastings algorithm. In the third section,
we discuss the modifications to the MCMC algorithm incorporated in
SCoPE to make it more efficient and economical. In the fourth section of the paper,
we provide illustrative results from our analysis of different cosmological
models with WMAP-9 and Planck data. Our work also provides a complete independent parameter estimation
analysis of the data using an independent MCMC code. The final section
is devoted to conclusions and discussions.

\section{Brief overview of Metropolis-Hastings algorithm}

The Metropolis-Hastings (MH) is one of the most widely used MCMC sampler,
in which the posterior i.e. $\pi(\theta)$ is sampled using a random
walk. A standard Markov Chain at each step $i$, randomly chooses
a candidate value $\theta_{i+1}$ from the proposal distribution $q(.|\theta_{i})$.
The candidate value only depends on the current data point $\theta_{i}.$
The new data point is then accepted with probability $\alpha=\min(1,\pi(\theta_{i+1})/\pi(\theta_{i}))$.
If the new data point rejected, the previous point is replicated
by increasing its weight by $+1$. The chain of data-points thus generated,
approximate the target posterior distribution $\pi(\theta)$. 

The proposal distribution is generally taken to be a Gaussian distribution
i.e.  $ q(\theta_{i+1}|\theta_{i})$ $=N\exp(-\frac{s}{2}u_{i}(C^{ij})^{-1}u_{j})$,
where $u_{i}=\theta_{i+1}-\theta_{i}$ and $C^{ij}$ is the covariance
matrix. $s$ is the step size. Theoretical optimum step size for an
ideal distribution that provide the best acceptance
rate is $s=2.4/\sqrt{n}$ for a $n$ dimensional MCMC sampler 
\cite{Haariro1999}. The covariance matrix
is provided as an input to the program. As the exact covariance matrix is unknown
before the analysis, in practice an approximate covariance matrix,
often based on some previous analysis, is provided. 
 % which is generally taken from some previous analysis.
If no prior information is available about the covariance between parameters then some
approximate diagonal matrix is also often used. However, in such cases
the acceptance rate of the sampler may reduce drastically
and can be ensured to remain reasonable only by trial and error. Therefore,
a better choice is to start with a initial guess diagonal covariance matrix 
and then to update the covariance matrix using the data points obtained so far
%to get
%few accepted data points. The covariance matrix can be adaptively
%updated using the data points obtained so far in a particular chain
from the chain \cite{Doran2004}.
This requires no prior knowledge about parameter covariance. 

Parallelization of MH sampler is generally done by running multiple chains. 
Whether it is better to run a longer chain than running
multiple short chains has been addressed and argued by many authors \cite{Geyer1992,Gelfand1990}.
%But in our analysis we run 
% problem as it takes long time. 
%If we run 
But, for running multiple
parallel chains proper mixing between the chains has to be
ensured. Therefore, each of the multiple chains has to be long enough so that
the it can represent an unbiased sample of the population.  Gelman-Rubin 
``$R$'' statistics \cite{Gelman1992} is generally used for
testing the mixing of chains. For convergence, the chains have to
be long enough such that $R$ is very close to unity. For practical
purposes it is taken as $R<1.2$. However, this criterion is often not sufficient
for ensuring proper sampling. 

In SCoPE multiple chains are used because running a single long chain in 
serial is computationally time consuming and hence is
not feasible for extensive problems of cosmological parameter estimation. Hence, it is desirable to devise an implementation of MH algorithm that allows the individual chains
to run in parallel and increase the acceptance rate of the models of a chain.
Apart from that the mixing of the chains are also necessary. 
%Along with that we must add something so that the chains mixes properly.
%Therefore, to accomplish there, we have made some modifications in the standard MH algorithm. 
The next section describes the modifications made to standard MH algorithm
to accomplish effective parallelization through prefetching together with 
all other above mentioned features, namely, enhanced acceptance,
regular covariance update from samples, as described in next section.

\section{Embellishing the standard Metropolis-Hastings algorithm}

\subsection{Prefetching}

%Due to the recent developments in high performance computing and availability
%of massively parallel systems, it is important to make sure that algorithms
%can run efficiently in parallel. In a MH algorithm the chains are
%intrinsically serial in nature that leads to long evolution time for a single chain. 
The MCMC can take advantage of parallel computing 
only by running number of distinct individual chains each on separate
processors
%by running more then one parallel chains 
as shown in \cite{Rosenthal2000}. However, the
drawback of this method is that the error related to the burn in steps
will be present in all the chains. Hence, the initial steps from all
the chains need to be removed, which leads to a huge wastage of computational
power. In certain problems, the time spent in the
burn-in phase may be significant if the convergence rate of the chains
is slow. More importantly, if the individual chains are not long enough then they
may not pick up the samples from same distribution due to clustering
within individual chains. 
Hence, the poor mixing of the chains is another major concern.
Apart from that a small chain may not sample the tail part of the
posterior adequately. % Therefore, a chain level parallelization may not
% be helpful to get a properly sampled posterior. 
Therefore, it is %often
extremely useful to speed up the generation of a single chain, through parallelization rather than using
multiple chains. When the state-space of the chain is high dimensional,
one possible way to do this is to divide the state-space into blocks,
and then handle each block
on a separate processor for each iteration of the Markov chain. 
This approach does indeed speed up generation
of a single chain, but requires additional effort, in carrying out
detailed analysis of the limiting distribution, in order to determine the appropriate
blocks. This may be difficult or even impossible in many cases, where
the conditional dependence structure in the limiting distribution
is complicated. Therefore, in our work we make the individual
chains parallel by precomputing several draws from the posterior distribution
ahead of time via multiple evolution of models simultaneously in parallel and
then use only the values that are needed \cite{Rosenthal2000}. 
%.The details of the method is discussed in 

\begin{figure}
\centering
\includegraphics[width=0.7\textwidth]{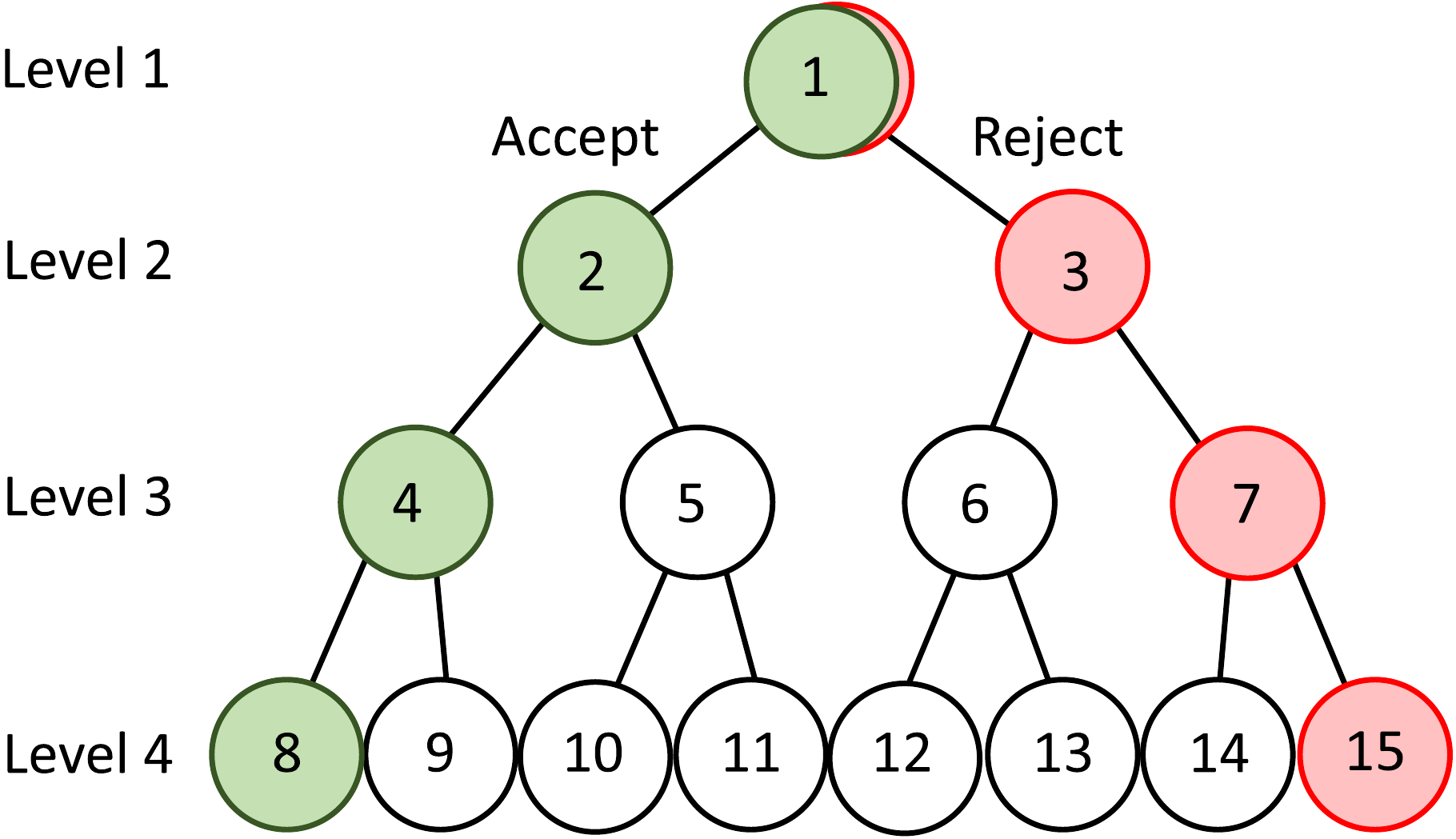}

\caption{\label{fig:presampling}Prefetching scheme is explained in the 
text with the help of above figure} 
\end{figure}

Prefetching is a draw level parallelization in a single chain \cite{Brockwell2006,Strid2009}.
The method can be explained by taking the binary tree of a Metropolis
algorithm as shown in Fig.~\ref{fig:presampling}. In a $k^{th}$
level binary tree there are total $2^{k}-1$ nodes, each of which
represents a possible future state of a metropolis algorithm. The
branches at the left child of any node represent the accepted steps and
the right child represents the rejected states. If we have enough
computational resources then all $2^{k}-1$ nodes can be evaluated
simultaneously and $k$ steps of a MCMC chain can be carried out in 
parallel simultaneously. 

Though the method of prefetching allows to parallelize a single chain,
it only uses $k$ steps out of $2^{k}-1$ computations. The rest of
the computations are not utilized. Therefore, many argue against
computing all the nodes of the binary tree. Rather if we know the
acceptance rate at a point of time from the previously accepted data
points, we can statistically identify and precompute only the most probable chain and
hence avoid any unnecessary wastage of computation power.
It is easy to see that if the acceptance probability at any point of
time is less than $0.5$ then the extreme right chain (1-3-7-15-..)
of Fig.~\ref{fig:presampling} will be most probable chain. In
a similar manner if the acceptance rate is more than $0.5$ then
the extreme left chain (1-2-4-8-..) will be the most probable chain.
Therefore, by pre-evoluting only the most probable chain, we parallelize
the code and at the same time we can manage the computational resources
in a better way. Hence, we have adapted this technique in SCoPE.

\subsection{Delayed rejection}

Delayed rejection is a concept proposed by A. Mira 
\cite{Mira2001,Tierney1999,Green2001,dram,Haario2006,Trias2009}.
One of the major problems with the MCMC method is the choice of the
step size for the proposal distribution. If the step size is large
then rejection rate increases because the variation of the likelihood
from sample to sample will be very large. On the other hand if the
step-size is taken to be very small then the convergence will be very
slow and the auto-correlation between samples will increase. So it
is important to choose the step-size optimally. The optimal step-size
can be chosen by trial and error method or by some other statistical
method, which we discuss in a later section. But even if we choose some
optimal step size for the proposal distribution, the acceptance rate
may not be very high. It will be better if the rejected sample from one step 
can be used to determine
the proposal distribution for the next sample. This increases acceptance
rate but at a cost of violation of the Markovian property. 
%As we are changing the proposal distribution from
%step to step therefore there should be some method to account for
%that change in the proposal distribution. 
But if we can find some method that can change the acceptance probability 
of the sample point to compensate the step size variation then that will be useful.
%One of such method was proposed
%by A. Mira \cite{Mira2001} and the method was named as Delayed rejection. 

The concept of delayed rejection can briefly be explained as follows.
Suppose at some step $i$, the position of a chain is $\theta_{i}=x$.
Suppose at this time a candidate $y_{1}$ is accepted from $q_{1}(x,y_{1})$
and accepted with probability

\begin{equation}
\alpha_{1}(x,y_{1})=\min\left(1,\frac{\pi(y_{1})q_{1}(y_{1},x)}{\pi(x)q_{1}(x,y_{1})}\right)
\end{equation}

\noindent as in the standard MH algorithm. For a Markov chain,
$q_{1}(y_{1},x)$ is time symmetric, i.e. $q_{1}(y_{1},x)=q_{1}(x,y_{1})$.
Therefore, the acceptance ratio only depends on the posterior. A rejection
at any step suggests that there is a local bad fit of the correct
proposal and a better one, $q_{2}(x,y_{1},y_{2})$, can be constructed
in light of this. But, in order to maintain the same stationary distribution
the acceptance probability of the new candidate, $y_{2}$, has to
be properly computed. A possible way to reach this goal is to impose
detailed balance separately at each stage and derive the acceptance
probability that preserves it. A Mira in \cite{Mira2001} has shown
that if the acceptance probability is taken as

\begin{equation}
\alpha_{2}(x,y_{1},y_{2})=\min\left(1,\frac{\pi(y_{2})q_{1}(y_{2},y_{1})q_{2}(y_{2},y_{1},x)\left[1-\alpha_{1}(y_{2},y_{1})\right]}{\pi(x)q_{1}(x,y_{1})q_{2}(x,y_{1},y_{2})\left[1-\alpha_{1}(x,y_{1})\right]}\right)\,,\label{eq:DR}
\end{equation}

\noindent then Markovian property of the chain will not get destroyed, but still
the sample choice can be made dependent on the previously accepted
data point. This particular procedure gives rise to a Markov chain
which is reversible with invariant distribution thus provides an asymptotically
unbiased estimate of the posterior distribution. 
%One of the advantages of
%having the return path go through $y_{1}$ is that since $y_{1}$
%has been rejected in the forward path, it means that $\alpha(x,y_{1})$
%is likely small. Since the term $1-\alpha_{1}(y_{2},y_{1})$ appears
%in the numerator of the acceptance probability the acceptance is higher
%at the second stage. Hence, the second sample is likely to get accepted
%with a higher probability. 

The delayed rejection method can be continued further, if the second
sample also gets rejected. A general acceptance probability for a $N$
step delayed rejection is proposed in \cite{Mira2001}. But for our
purpose we have just consider a $1$ step delayed rejection algorithm.
In our algorithm if a data point at a particular step is rejected then we decrease
the step size and sample a new data point. The acceptance probability
for the new data point is calculated using Eq(\ref{eq:DR}). This
increases the acceptance rate of a chain and decreases the autocorrelation
between the data points by keeping the chain in motion instead of
getting stuck at some step.

\begin{table}
\centering
\begin{tabular}{|c|c|c|c|c|c|}
\hline 
~~~ No DR \& ~~~ & ~~~ DR \& ~~~& DR \& & DR \& & DR \& & DR \&\tabularnewline
~~~ No PF ~~~ & ~~~ No PF ~~~& ~~~1 PF ~~~& ~~~2 PF~~~ & ~~~3 PF~~~ & ~~~4 PF~~~\tabularnewline
\hline 
\hline 
24.53\% & 49.12\% & 67.37\% & 84.71\% & 92.79\% & 97.37\%\tabularnewline
\hline 
\end{tabular}

\caption{\label{tab:DRPF}Acceptance rate of a chain with and without delayed
rejection (DR) and pre-fetching (PF) for a particular run of the MCMC
code. Counts exclude the samples from the burn in steps. }
\end{table}

Table (\ref{tab:DRPF}) lists the acceptance rate of a chain for
a SCoPE run. The acceptance rate in the initial steps of the burn
in process is poor as the covariance matrix is not known properly.
However, as the covariance matrix gets updated due to the adaptive
covariance update, the acceptance rate gradually increases to some
fixed value. Therefore, the initial burn in steps are not included
in the acceptance rate analysis. Without any delayed rejection and
prefetching the normal acceptance rate is as poor as $~25\%
$. This means most of the steps get replicated several times. There
are many steps that get replicated more than 20 times, increasing
the autocorrelation of a chain. With delayed rejection the acceptance
rate increases to $~50\%$. For the pre-fetching the acceptance rate
is defined as (Number of accepted data point)/(Number of steps). In
a $n$ step pre-fetching we are running $n$ parallel computation
in each step. Using only $3$ to $4$ prefetching steps we are able
to reach more than $90\%$ acceptance rate.

\subsection{Inter-chain covariance adaptation}

The practical problem in implementing MH is the tuning problem of
the proposal distribution $q$ so that the sampling is efficient.
%If the step-size for generating the next sample at any stage is taken
%to be large then most of the samples will get rejected and hence the
%chains will be less efficient. Also if we take the chain size to be
%very small then acceptance rate will definitely improve, but the clustering
%of the chains will increase and hence the chains may not sample the
%entire target distribution properly. 
One of the recent improvements
in the MCMC efficiency is to introduce adaptive samplers. The adaptive
MCMC uses the sample history and automatically tune the proposal distribution
in the sampling process e.g.\cite{Gilks1998,Haariro2001,Roberts2007,Andrieu2006}.
In adaptive metropolis algorithm \cite{Haariro2001,Solonen2012},
the covariance matrix from the samples obtained so far is
used as the covariance of a Gaussian proposal. Hence the new candidates
are proposed as $\theta_{i+1}\sim N(\theta_{i},\Sigma_{i})$ where
$\Sigma_{i}=\rm Cov(\theta_{1,...,}\theta_{i})+\epsilon I$ and $\rm I$ is
the identity matrix. The method can be used to make the MCMC algorithm
adaptive. 

The most common parallelization scheme of MCMC method is to run parallel
chains instead of running a single one. 
%There are many papers where
%people argue about running single chain in stead of multiple chains,
%however in our paper we run multiple chains, because running a single
%chain is not feasible for a cosmological parameter estimation problem
%due to time constraints. 
If in each chains proposal distribution
is adapted using the local covariance matrix then the acceptance probability
%of each data points in
a chain may improve, however, the inter-chain
mixing will not improve. If some chain stuck at some local minima
then the local covariance matrix corresponding to that chain will
be erroneous. 
%So in case of the local covariance update each chain
%may sample some local peak distribution and may not explore the entire
%target distribution. 
So, in case of a local peak the local covariance update will give covariance
corresponding to the local peak.
In that case, the mixing of chains will slow down
and sampling may not be proper. Therefore, in this paper we have adapted
the concept of the inter-chain covariance update in the adaptation
technique. We run several parallel chains, and randomly update
the covariance matrix taking the data points accepted till then from
all the chains. This means we have used the covariance as $\Sigma_{i}=\rm Cov(\theta_{1_{1}}..,\theta_{i_{1},}\theta_{1_{2},..,}\theta_{i_{2}....}\theta_{1_{n}},...,\theta_{i_{n}})$,
where, $n$ is the number of chains. This inter-chain covariance adaptation
\cite{Solonen2012} speeds up the mixing of the the chains and covers
the sample space faster. 

\begin{figure}
\centering
\includegraphics[trim=1cm 0cm 0cm 0cm, clip=true, width=1.1\textwidth]{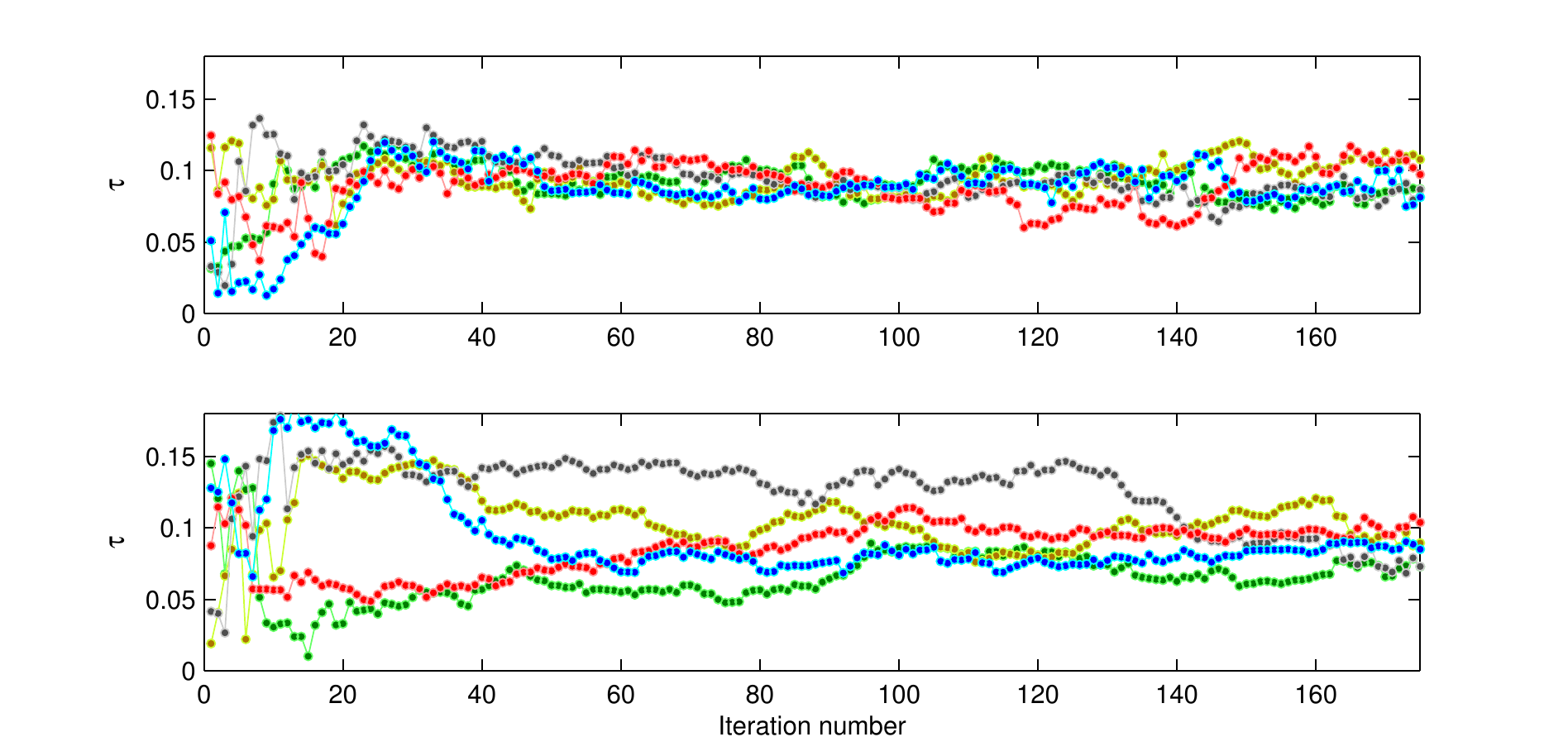}

\caption{\label{fig:convergence}The plots show the value of the parameter
$\tau$ for first 175 steps of the accepted MCMC chain on WMAP-9 year
data. The top plot shows the five chains from a MCMC run where
inter-chain adaptive covariance update is applied and the bottom plot
show chains from a run without any covariance update and step-size update.
It can be seen that for the case of adaptive covariance and step-size
update almost all the chains converge within first 20-30 accepted
steps, however when adaptive covariance update is not applied, the
chains take really long time to converge (approximately 160 steps).
The choice of parameter $\tau$ in above is inconsequential. Similar improvement is
seen for any other parameter.}
\end{figure}

The value of the covariance matrix will freeze after few adaptations
and hence we will be using same Gaussian proposal after few steps, which
is important to guarantee proper sampling. The inter-chain covariance
update speeds up the mixing of the chains and thus the Gelman Rubin
statistics converges within very few steps. However, if the adaptive
covariance is not frozen, it may give rise to unfair sampling as the
Gaussian proposal will vary between steps. Therefore, the process
of the adaptive covariance calculation is only used for the initial
burn in process and after that the adaptation is stopped, during which
the covariance calculation attains partial convergence. The effectiveness
of the process can be seen from Fig.~\ref{fig:convergence}. 

In the Fig.~\ref{fig:convergence} we have shown first 175 values
of the quantity $\tau$ for five chains of a MCMC run. It can be seen
that when in inter-chain adaptive covariance update is incorporated,
the chains converge just after 20-30 steps, which is 
really fast. %The chains just converge after 20-30 iteration, 
Whereas in the case where no inter-chain covariance update
is incorporated, the chains take more than 150 steps to attain
convergence. Therefore, the burn in steps significantly reduce.
%and we need to remove lesser number of burn in steps in this case. 

\section{WMAP-9 and Planck parameter estimation with SCoPE}
\label{sec:6param}
In this section we show examples of cosmological parameter
estimation with SCoPE using WMAP-9 and Planck data using 
the likelihood estimators provided by the respective teams. The first example is
for the standard 6 parameter $\Lambda$CDM model. Here we have shown
a comparative analysis of the WMAP-9 and Planck results. 

Standard $\Lambda$CDM model parameter estimation from WMAP-9 and Planck data sets
has been carried out by many authors. The 6 main parameters which
are used for standard $\Lambda$CDM parameter estimation are physical
baryon density ($\Omega_{b}h^{2}$), physical matter density ($\Omega_{m}h^{2}$),
Hubble parameter ($h$), reionization optical depth ($\tau$), scalar
spectral index ($n_{s}$), amplitude of the temperature fluctuations
($A_{s}$). 

\begin{figure}
\centering
\includegraphics[trim=0cm 6cm 0cm 6cm, clip=true, width=1.0\textwidth]{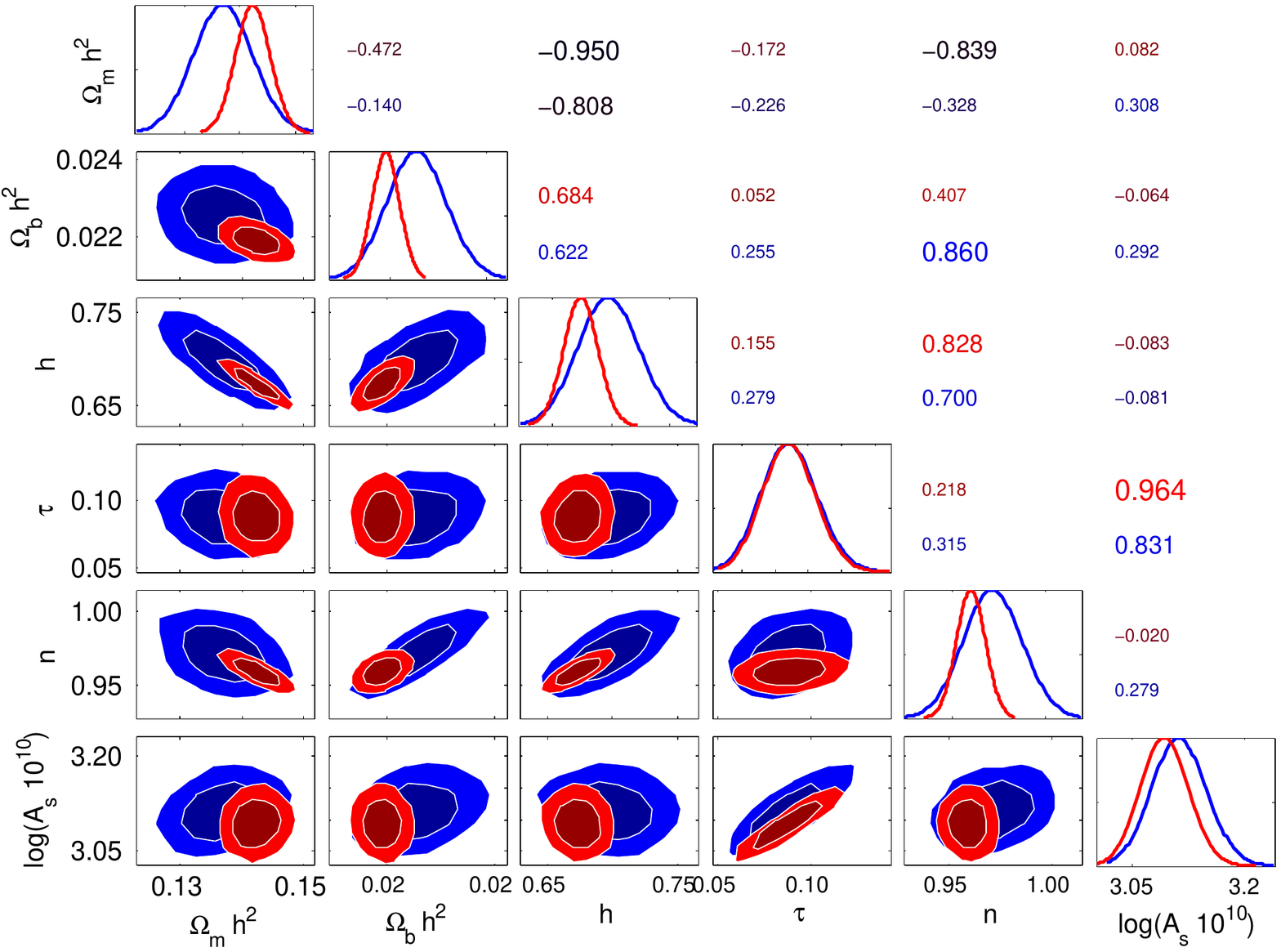}

\includegraphics[trim=0.8cm 0cm 0.8cm 0.5cm, clip=true, width=1.0\textwidth]{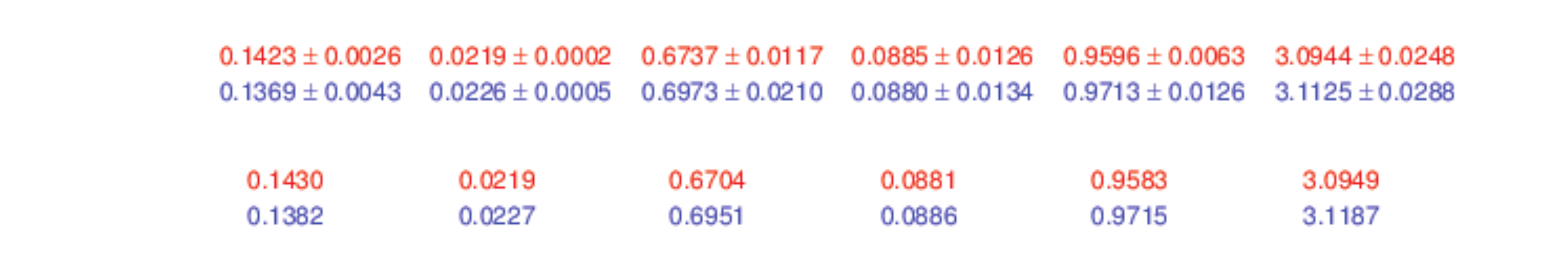}

\caption{\label{fig:standardLCDMmodel}Results of cosmological parameter estimation from
WMAP-9 (in blue) and Planck (in red) for the standard 6 parametric
$\Lambda$CDM model. The lower triangle panels show plots of the $68\%$ and $95\%$ confidence
conturs for pairs of parameters. The upper triangle mention the covariance
between the pairs parameters. The diagonal plots are the 1 dimensional marginalized
distribution of the parameters. The average, standard deviation and
the best fit values of these parameters are tabulated below the panels. }
\end{figure}

The result from WMAP-9 and Planck simulations are shown in Fig.~\ref{fig:standardLCDMmodel}. 
Likelihoods are calculated using
the likelihood software provided by WMAP-9 and Planck team \cite{WMAP9,Planck}.
For calculating the Planck likelihood, we have used low like\_v222.clik
and CAMspec\_v6.2TN\_2013\_02\_26.clik and added them up. We have
not used the actspt\_2013\_01.clik data set as that is only used to
obtain constraint very high multipoles of the CMB power spectrum. 
We have used only the standard 6 parameter
model. All the nuisance parameters are fixed to their average values
from Planck+WP+highl+BAO parameter estimation as given in \cite{Ade2013}.
The result of cosmological parameters allowing variation in the nuisance parameters is shown
in a later section of the paper. 

The result from WMAP-9 and Planck for the standard $6$ parameter model is 
shown in Fig.~\ref{fig:standardLCDMmodel}. It can be seen that the error bars on 
the parameters decreases substantially for Planck. The results from our analysis 
matches very well with the results quoted in Planck papers \cite{Ade2013}. 
The small deviations are due to the fact that we have fixed the nuisance parameters 
their average values. The full analysis result with all the nuisance parameters 
is shown in a later section.  

\subsection{Different dark energy parametrization}

Recent data from Planck suggest that the power at the low multipoles
of the CMB power spectrum is lower than the theoretically expected
value. In \cite{Das2013,Das2013a}, it is shown the low power
at the low multipoles can originate from the ISW effect, which can
only affect the CMB low multipoles without introducing any significant features
in any other observables. ISW effect comes from the late time expansion
history of the universe, which is controlled by the properties of
Dark energy in the universe. Therefore, it is important to check if
standard variants of dark energy models provide a better fit to the CMB
data. 

\begin{figure}
\centering
\includegraphics[trim=0.54cm 0cm 0.54cm 0cm, clip=true, width=1.0\textwidth]{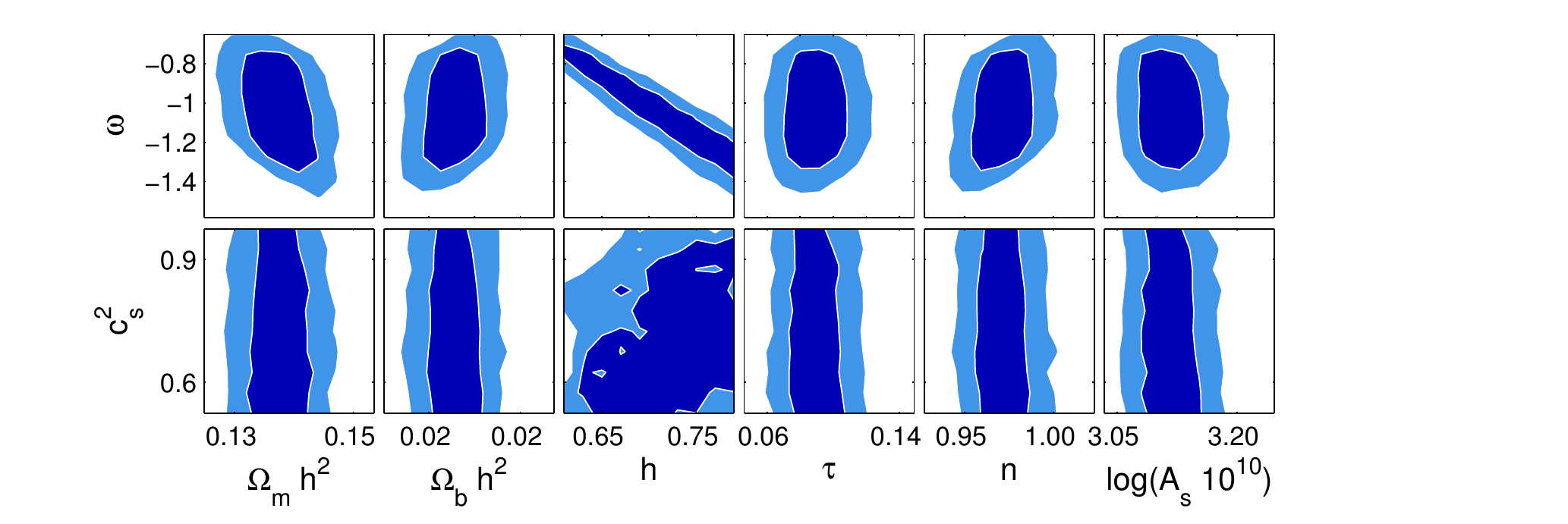}

\includegraphics[trim=0cm 11cm 0cm 11cm, clip=true, width=1.0\textwidth]{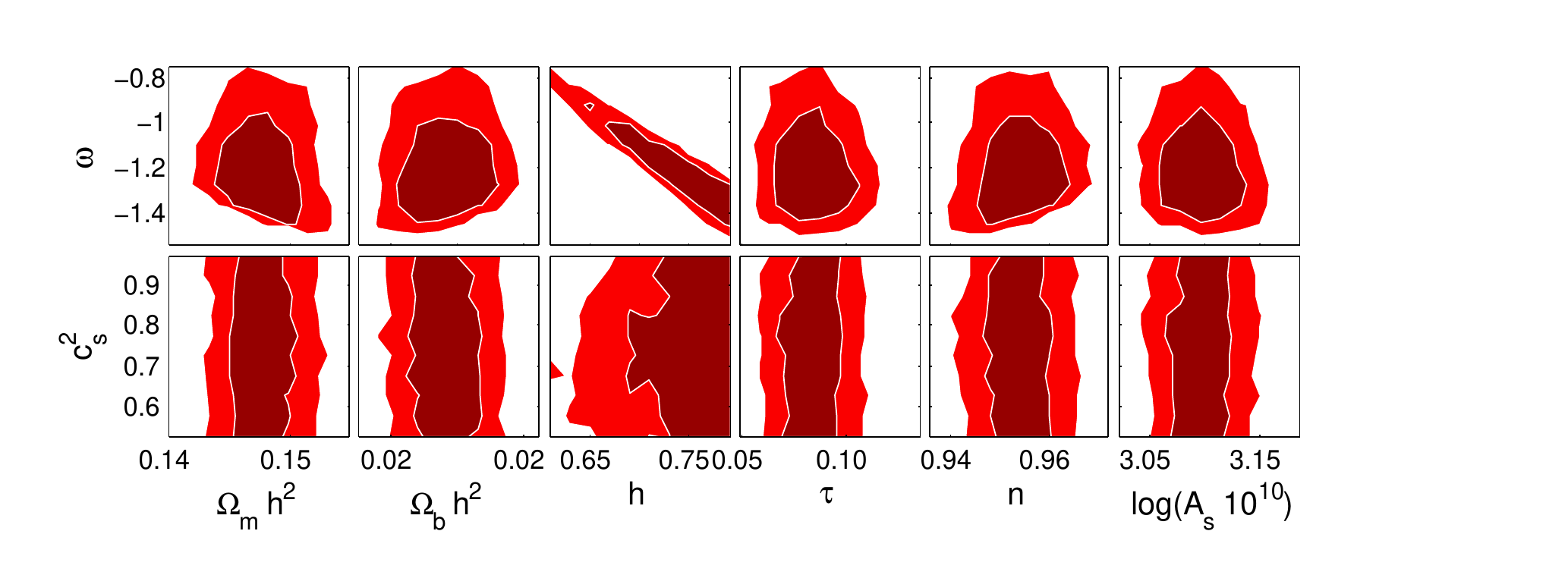}

\caption{\label{fig:2DconstantEOS}The $68\%$ and $95\%$ confidence contour for
the constant equation of state dark energy model. The blue plots are 
form the WMAP-9 results and the red plots are from Planck results. The dark energy
sound speed $c_{s}^{2}$ is almost flat which shows that $c_{s}^{2}$
cannot be constrained using WMAP-9 or Planck results.}
\end{figure}

\begin{figure}
\centering
\includegraphics[trim=2cm 11cm 2cm 11cm, clip=true, width=0.8\textwidth]{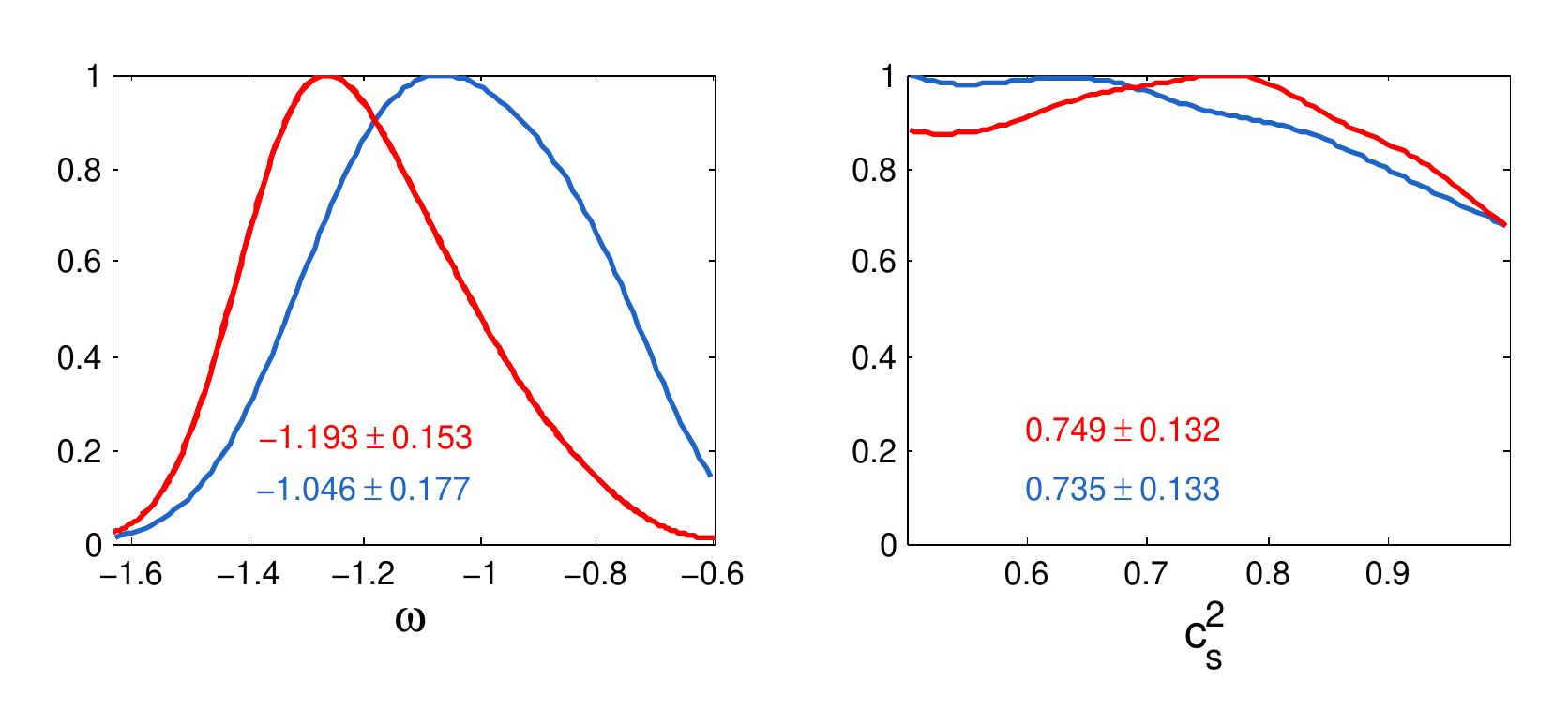}

\caption{\label{fig:1DconstantDE}One dimensional likelihood plot for the constant
equation of state dark energy. Blue plot is showing the WMAP-9 result and the 
red curve show is for the Planck data.The plots shows that the peak of $\omega$
is very much close to the $-1$ for WMAP-9. However for Planck results the peak has
shifted to $\sim -1.2$. The plots shows that $c_{s}^{2}$
cannot be constrained by the data set. The likelihood curve for $c_{s}^{2}$
is almost flat.}
\end{figure}

Several dark energy models are available in literature such
as cosmological constant, quintessence \cite{Ratra1988,Turner1997,Scherrer2008},
k-essence \cite{Scherrer2004,Chimento2004,Chiba2002}, phantom fields
\cite{Caldwell2002}, tachyons \cite{Bagla2003} etc. Different
empirical parameterizations for the dark energy are also proposed by
authors. Here we have tested some of the standard dark
energy models. The generalized equation for a fluid assumption of
dark energy perturbation is shown in \cite{Das2013}. There are two
parameters for quantifying dark energy perturbations, which are
the equation of state $\omega$ and the dark energy sound speed $c_{s}^{2}$.
There are models where the $\omega$ is a function of scale factor. We analyze 
two models and try to fix these parameters using SCoPE.

\begin{figure}
\centering
\includegraphics[trim=3.6cm 10cm 3.9cm 10cm, clip=true, width=0.46\textwidth]{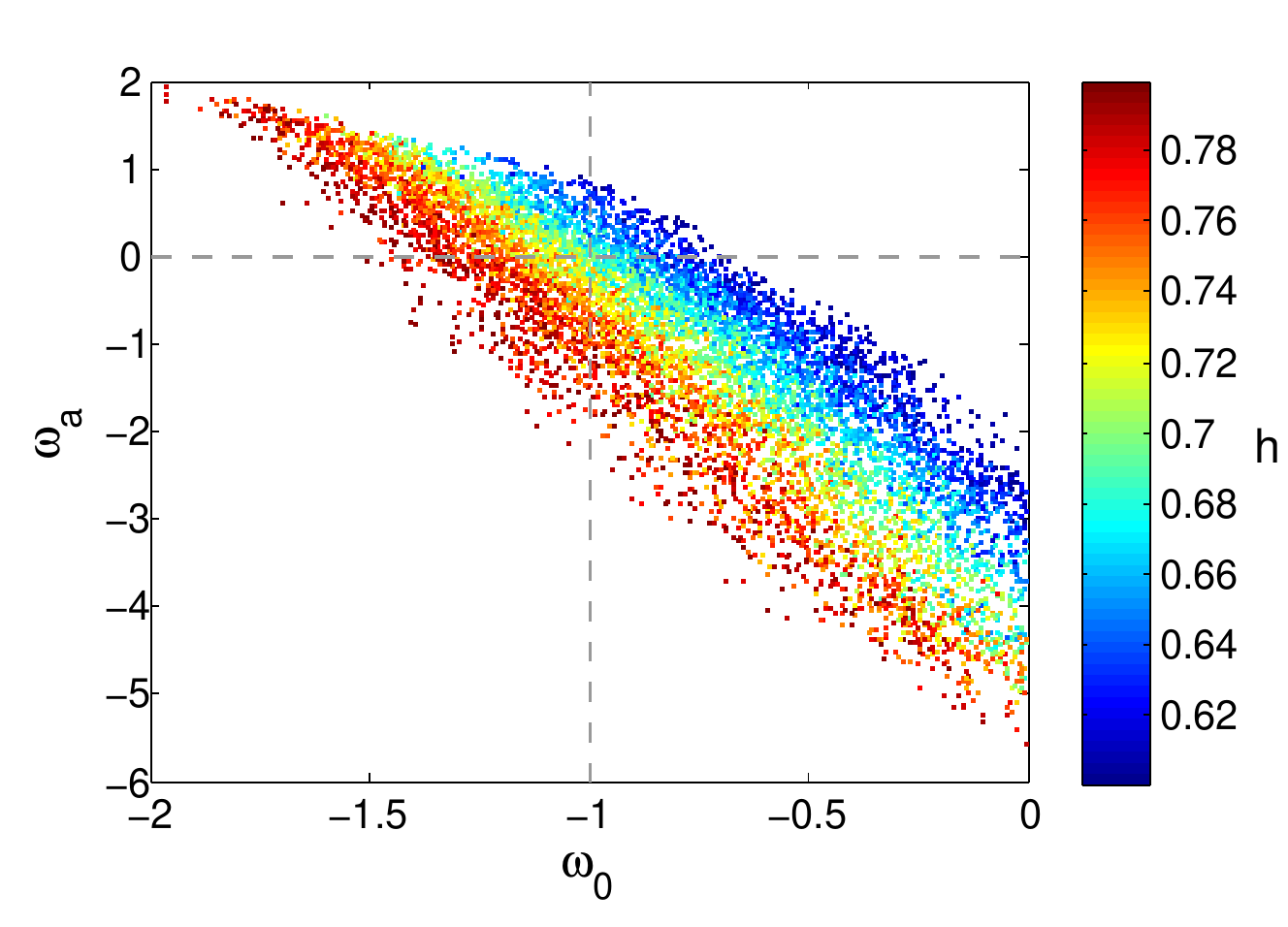}
\includegraphics[trim=3.6cm 10cm 3.9cm 10cm, clip=true, width=0.46\textwidth]{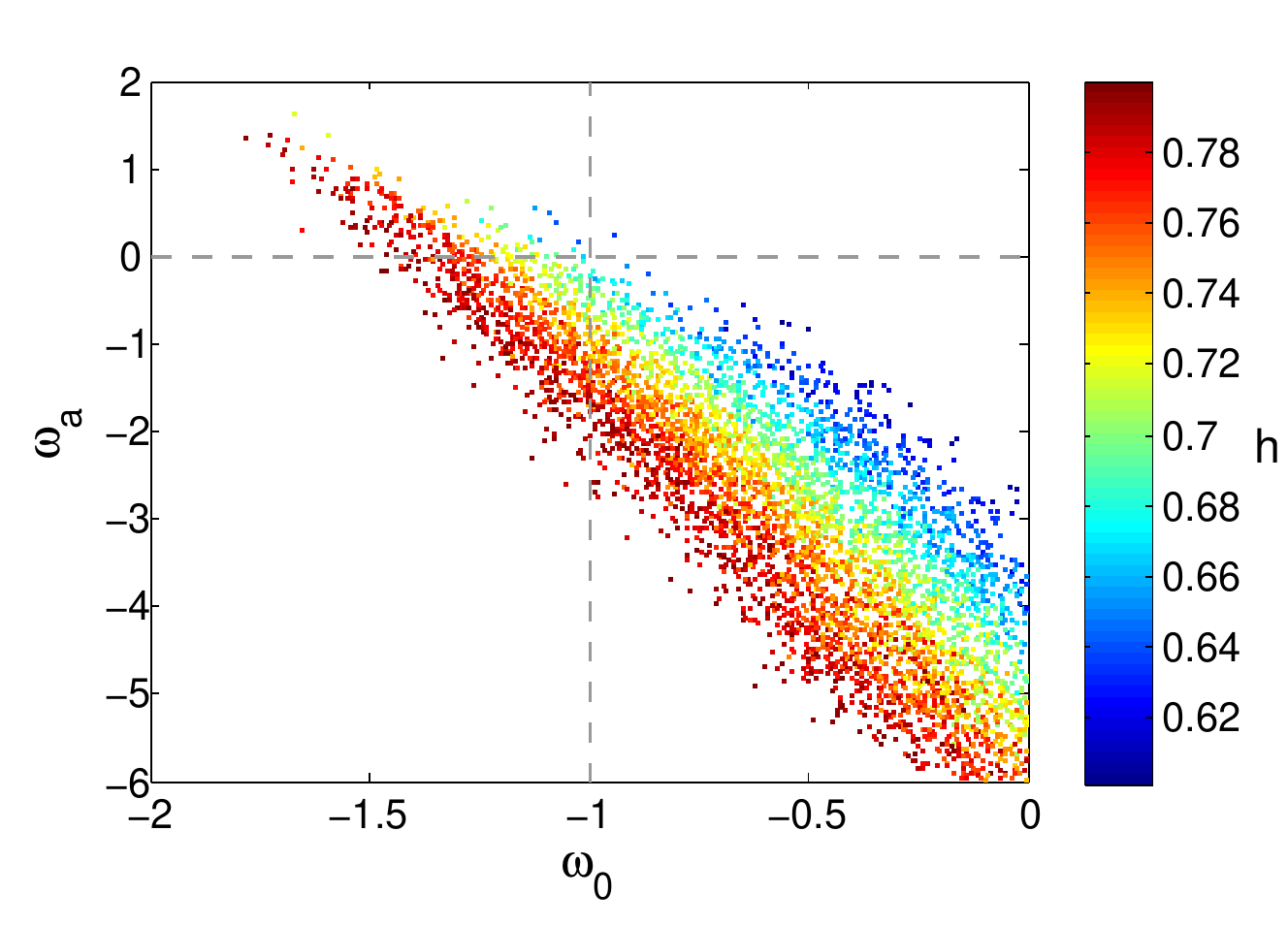}
\caption{\label{fig:CPLwwh}Plot shows scatter distribution of the samples $\omega_{0}$
vs $\omega_{a}$ plane with the value of $h$ color coded on it. It is clear that $\omega_{0}$ and
$\omega_{a}$ are strongly negatively correlated. The black dotted
lines shows the $\Lambda$CDM model. The left plot is for WMAP-9 year data set
and the right plot is for Planck data set. Though for WMAP-9 data the $\Lambda$CDM 
model is at the center of the distribution, the Planck data shows slight deviation 
from the $\Lambda$CDM model. The $\Lambda$CDM model is located at the edge of the 
distribution for Planck data.}
\end{figure}

\paragraph{Constant equation of state dark energy model}

For the constant equation of state dark energy, $\omega$ is constant 
and hence %there are two parameters
%for constraining the dark model, the parameters that 
we need to fix %are 
$c_s^2$ and $\omega$ along with the 
%energy perturbations. One is the equations
%of state of dark energy $\omega$and the other is the dark energy
%sound speed $c_{s}^{2}$ (square of the sound speed). Apart from that
%there will be 
other 6 standard model cosmological parameters. We 
run SCoPE for a 8 parameter model.
The covariance between the standard model parameters is almost similar
to that of the standard model parameters. Therefore, we do not show
the plots of those parameters. 
%Using these simulations we try to constraint
%the $\omega$ and $c_{s}^{2}$ of the dark energy. 
The $68\%$ and $95\%$ confidence contours of $\omega$ and $c_{s}^{2}$ with other 6 standard
model parameters are shown in Fig.~\ref{fig:2DconstantEOS}. It
can be seen that the dark energy equation of state is strongly negatively
correlated with Hubble parameter. This strong correlation is expected
as dark energy equation of state $\omega$ changes the expansion history
of the universe that leads to the change in %is changing and the 
distance of the last scattering
surface. % is varying. %Therefore, to compensate that distance 
This change is actually compensated by the change in 
Hubble parameter. % If this change of distance of the
%last scattering surface is not accounted for then the power spectrum
%will shift horizontally towards higher or lower $l$'s. 
 Apart from $H$, dark energy equation of state is almost uncorrelated with any
other standard model parameter. The second row shows that
 $c_{s}^{2}$ is almost uncorrelated  with any other standard model parameters.
The data from WMAP-9 or Planck are not good enough to
put any constraint on $c_{s}^{2}$. The dark energy sound speed
mainly affects the low multipoles of the CMB power spectrum. However, the
effect is not strong enough to put any bound on sound speed. In 
Fig.~\ref{fig:1DconstantDE} we show the one dimensional marginal probability
for $\omega$, $c_{s}^{2}$. It can be seen that the dark energy
equation of state ($\omega$) peaked near $-1$ for WMAP-9, indicating that the
standard $\Lambda$CDM model is very good assumption for the dark
energy model. Though the peak shifted towards $\omega \sim -1.25$ for the Planck data, 
which mainly caused by the power deficiency at the low multipoles of $C_l^{TT}$. 
Also, the probability distribution for $c_{s}^{2}$ is almost
flat. Therefore, we can conclude that it is almost impossible to 
put any constraint on $c_{s}^{2}$ using
the present CMB data sets.

\begin{figure}
\centering
\includegraphics[width=0.99\textwidth]{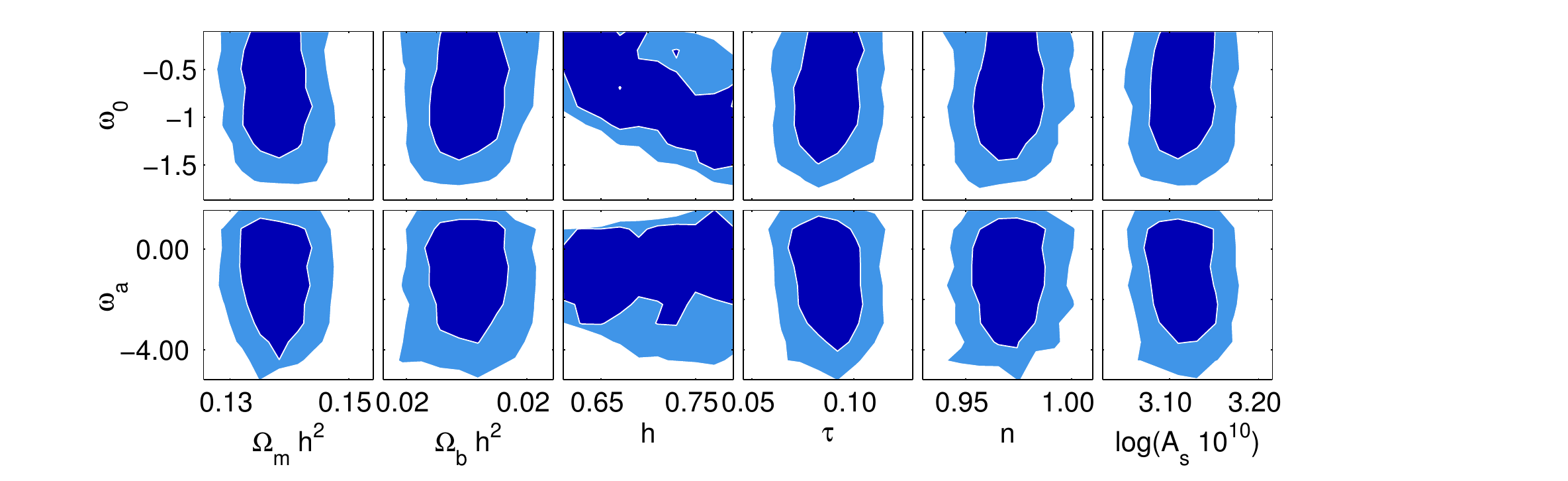}
\includegraphics[trim=2.1cm 0.0cm 0.0cm 0.0cm, clip=true, width=0.80\textwidth]{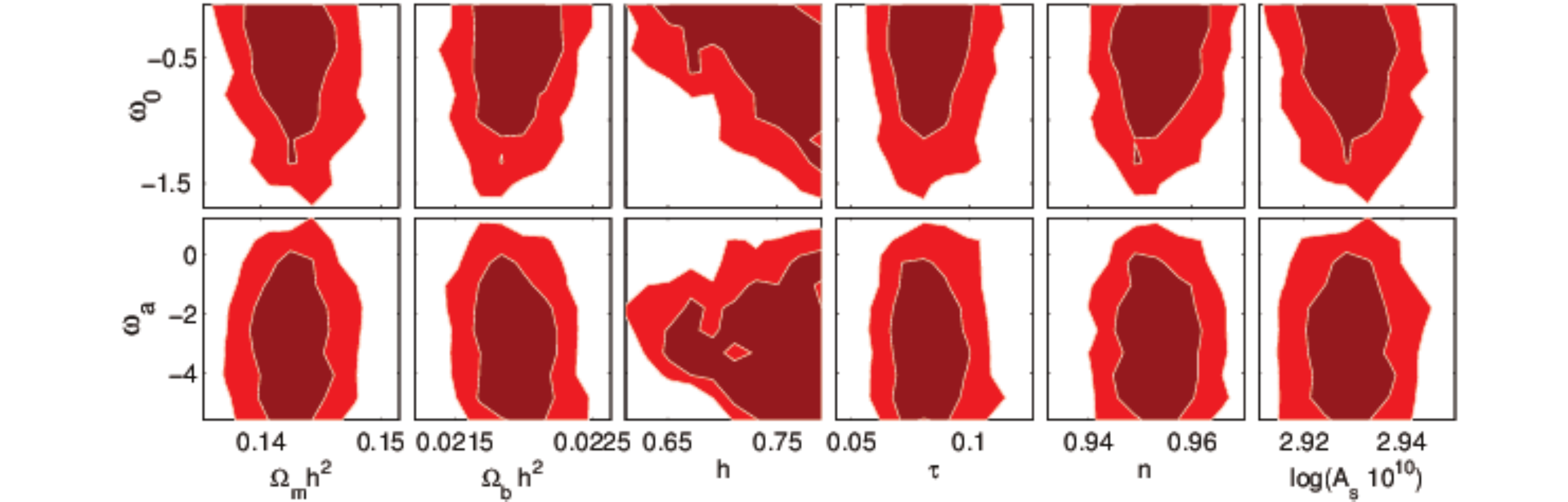}
\caption{\label{fig:CPL2D}2 dimensional likelihood distributions of $\omega_{0}$
and $\omega_{a}$ with other $6$ standard model parameters. The blue plots are for WMAP-9
data and the red plots are for the Planck data. Plots shows that $\omega_0$ and 
$\omega_a$ are almost uncorrelated with all the standard model parameters except $h$.}
\end{figure}

\begin{figure}
\centering
\includegraphics[trim=0.0cm 10cm 0.0cm 10cm, clip=true, width=0.90\textwidth]{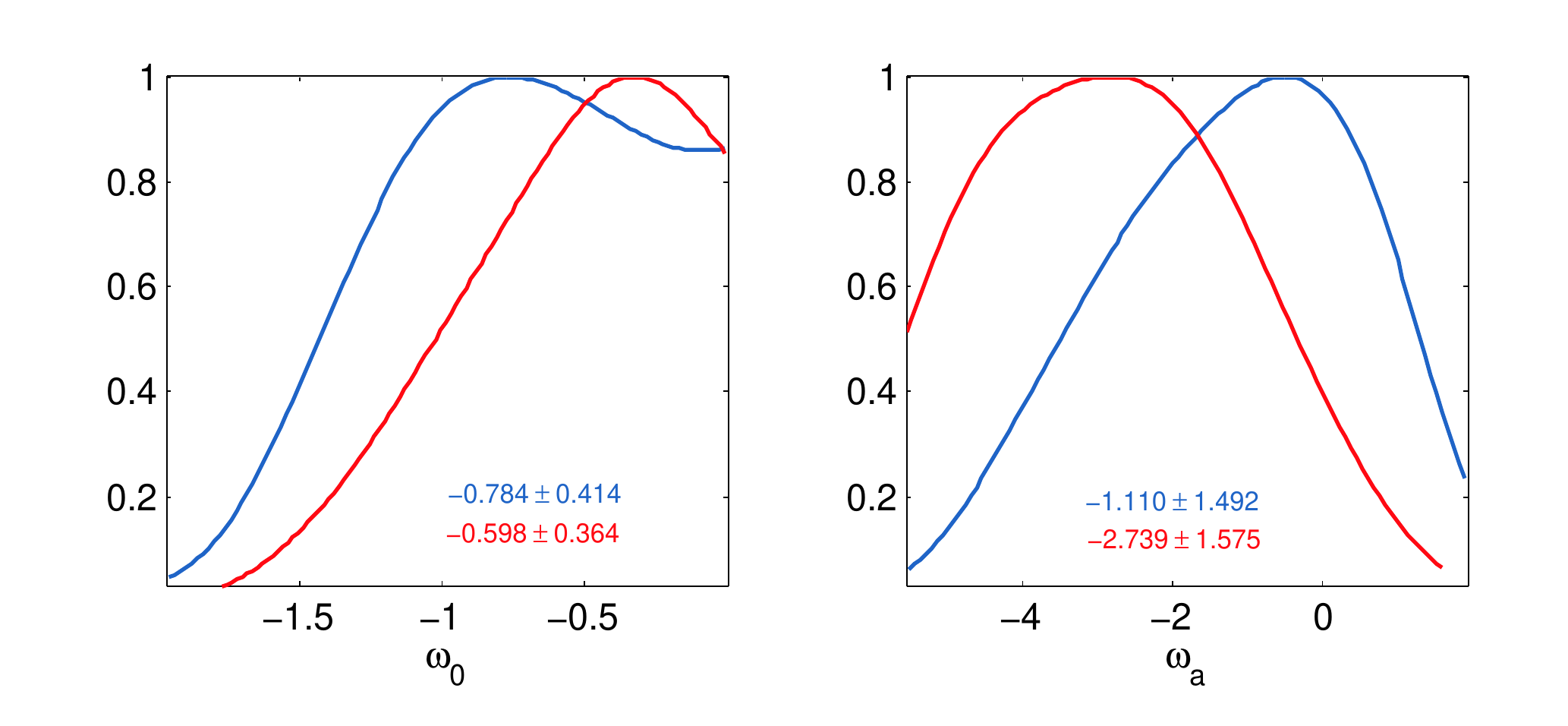}

\caption{\label{fig:CPL1D}1-dimensional marginalized probability distributions
of $\omega_{0}$ and $\omega_{a}$. Blue curve is for WMAP9 and Red curve is corresponds to 
Planck data. The plots show that the upper bound on  $\omega_{0}$ is not strongly 
constrained by the data set.}
\end{figure}

\paragraph{CPL Dark energy parametrization}

The CPL dark energy parametrization is an empirical dark energy parametrization,
introduced by Chevallier and Polarski \cite{Chevallier2001} and later
by Linder \cite{Linder2003}. In the CPL dark energy model the equation
of state of the dark energy is taken as 

\begin{equation}
\omega(a)=\omega_{0}+\omega_{a}(1-a)\,.
\end{equation}

%It is an empirical dark energy model and have not motivated by any
%theoretical model. If omega varies slowly then we can expand any equation
%of state of that dark energy model in terms of the Teller expansion
%up to the linear term and construct the equation of state for that
%dark energy. 
In the analysis we try to estimate $\omega_{0}$ and $\omega_{a}$ along 
with other $6$ standard model parameters. We have taken the $c_{s}^{2}=1$.
In Fig.~\ref{fig:CPLwwh} we have plotted $\omega_{0}$ vs $\omega_{a}$
in the scatter diagram and we have color coded $h$ in it. It can see
that there is a negative correlation between $\omega_{0}$ and $\omega_{a}$.
In Fig.~\ref{fig:CPL2D} we have plotted the two dimensional likelihood
distributions. It shows that there are strong negative correlation
between $\omega_{0}$ and $h$. In Fig.~\ref{fig:CPL1D} we have
shown the one dimensional marginal probability distribution of the
dark energy parameters i.e. $\omega_{0}$ and $\omega_{a}$. It can
be seen that upper bound on $\omega_{0}$ is not tight. Also the results from 
WMAP-9 and Planck differ significantly from one another. 

\begin{figure}
\centering
\includegraphics[trim=1.0cm 7.0cm 0.0cm 7.0cm, clip=true, width=0.49\textwidth]{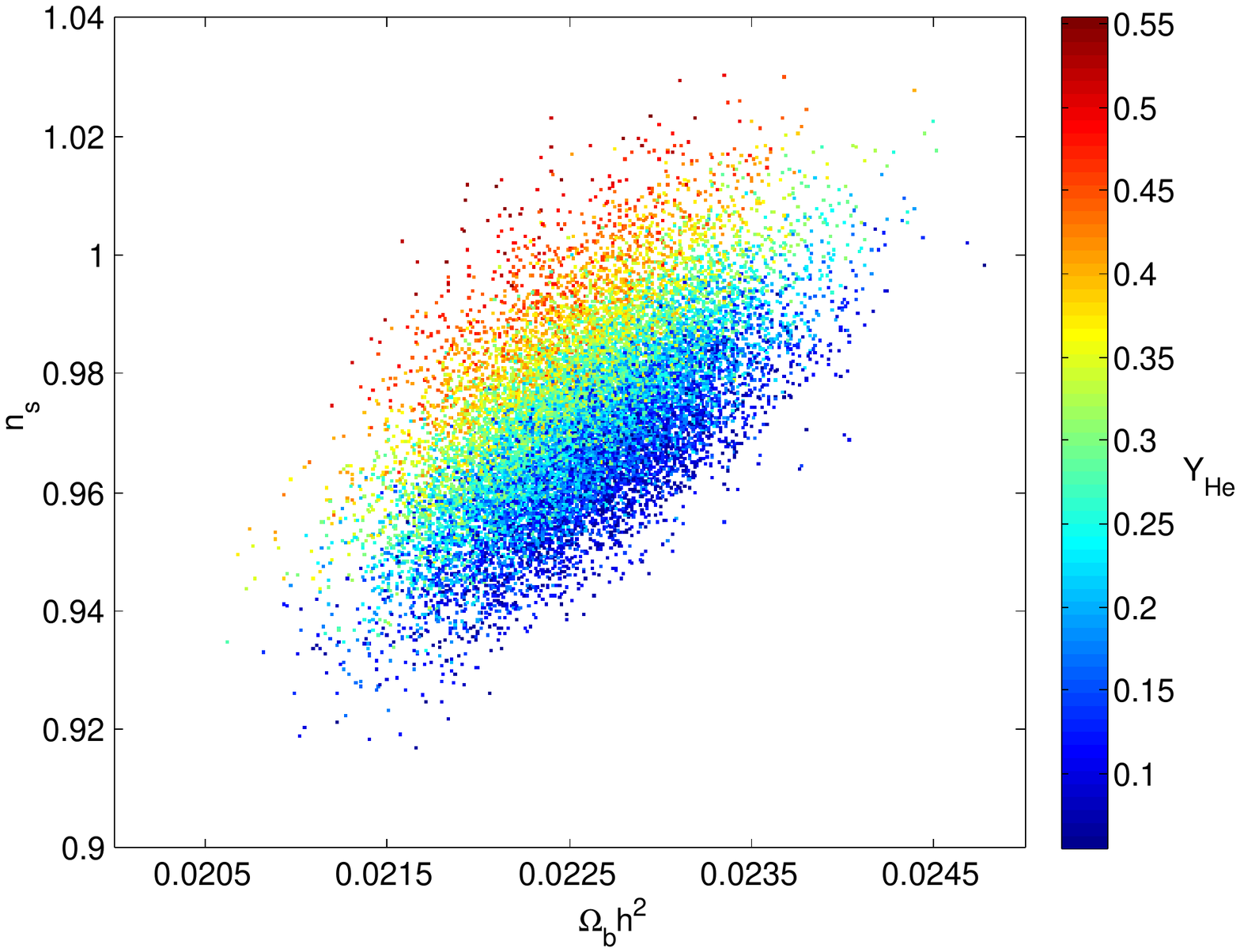}
\includegraphics[trim=1.0cm 7.0cm 0.0cm 7.0cm, clip=true, width=0.49\textwidth]{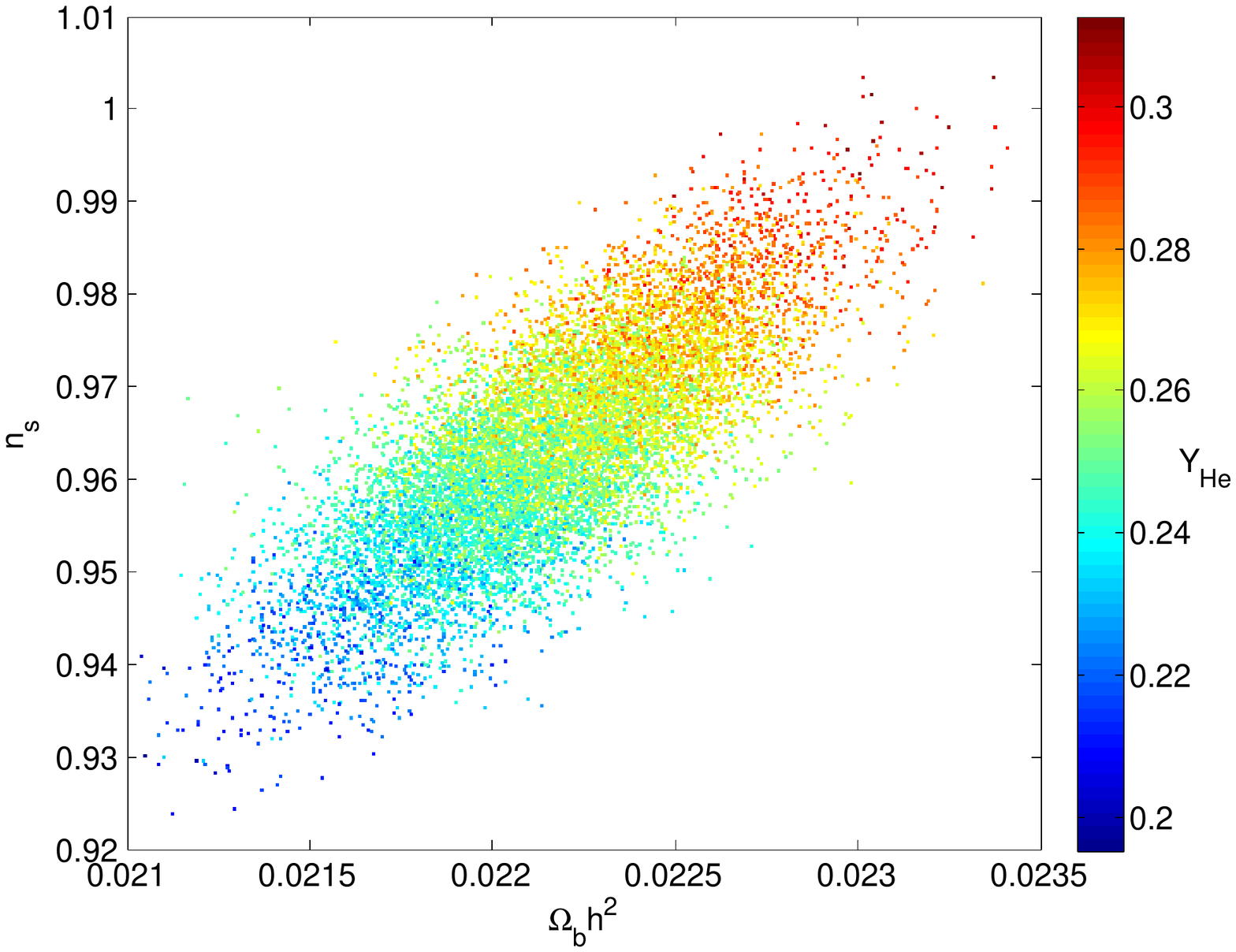}
\caption{\label{fig:msoby}Scattered plot of $n_s$ vs $\Omega_b h^2$ 
color coded with $Y_{He}$. The plots show that if $Y_{He}$ is increased
$n_s$ increases and $\Omega_b h^2$ decreases. The left plot is for 
WMAP-9 year dataset and right plot is for the Planck dataset.}
\end{figure}

\subsection{Helium fraction}

\begin{figure}
\centering
\includegraphics[trim=1.0cm 10.2cm 0.0cm 10.7cm, clip=true, width=0.89\textwidth]{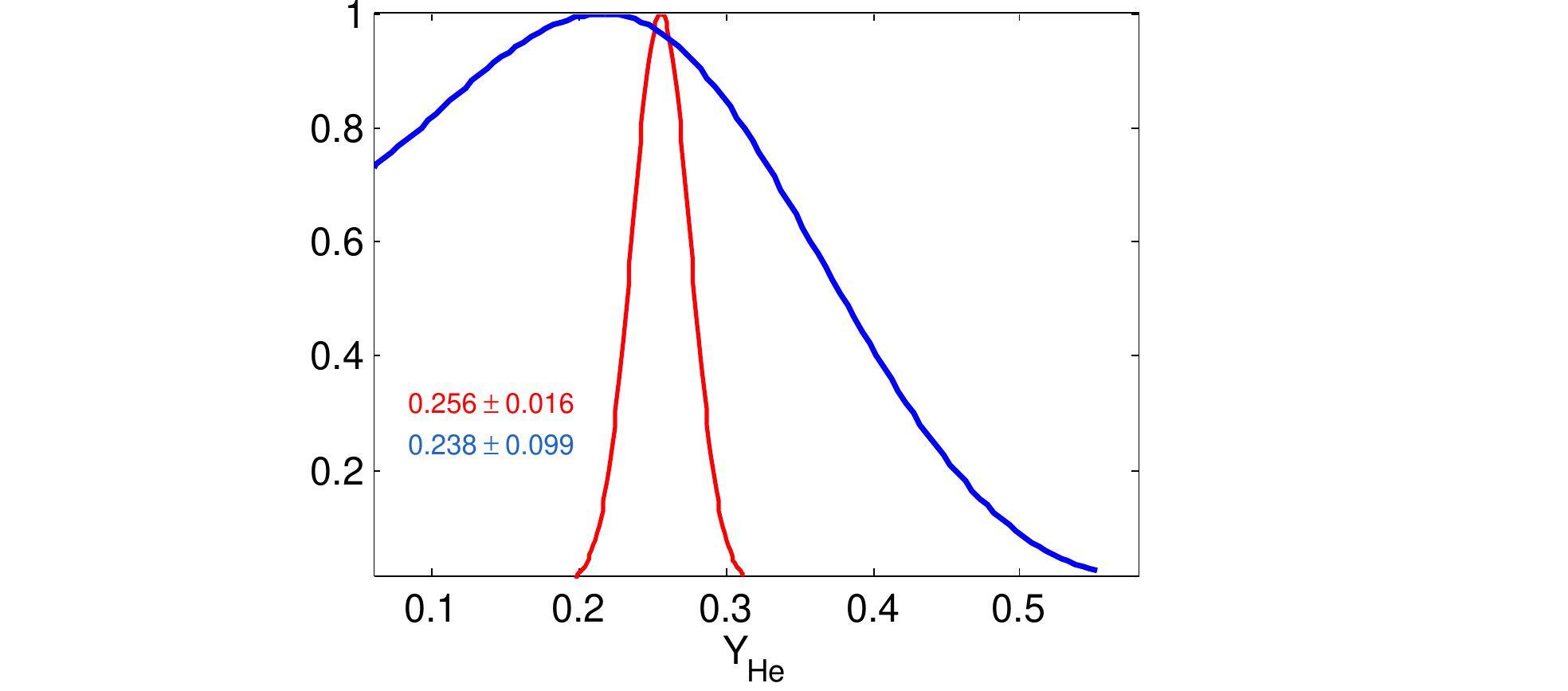}
\caption{\label{fig:heliumlikelihood}Plot shows 1 dimensional marginal probability
distribution for $Y_{He}$. Blue plot is for WMAP-9 year data and Red plot is for Planck 
data. The plots shows that the constrain on $Y_{He}$ highly improved by the Planck 
observation relative to WMAP.}
\end{figure}

Though the primordial helium fraction, $Y_{He}$ does not affect the CMB %power spectrum 
perturbations directly, %it is not included in the standard 6
%parameter model. But $Y_{He}$ 
its indirect effect on recombination and reionization can change the CMB power spectrum \cite{Ade2013,Trotta2004}.
For understanding the effects of $Y_{He}$ on the CMB power spectrum
we can use the free electron fraction, $f_{e}=n_{e}/n_{b}$,
where $n_{e}$ is the free electron number density and $n_{b}$ is
the baryon number density. Before HeII recombination ($z>6000$) all
the electrons were free and hence free electron fraction was $f_{e}=1-Y_{He}/2$.
After HeII recombination the free electron fraction drops down to
$f_{e}=1-Y_{He}$, and this remains valid up to redshift $z=1100$. Then
hydrogen and HeI recombines and hence $f_{e}$ drops down to almost
to values close to zero. Finally, after reionization at late time $f_{e}$ becomes $1-Y_{He}$.
Therefore, if $Y_{He}$ is changed then the free electron fraction
will change at various epochs. This will lead to the change in recombination
and reionization redshift. It may appear that if Helium fraction is changed then to compensate
it we can change the baryon fraction in the universe. 
However, $Y_{He}$ does not change the ratio of the
even and odd peaks. Therefore, changing baryon fraction does not actually
compensate the features induced from $Y_{He}$. An increase in $Y_{He}$
supress the power of the CMB power spectrum at the high multipoles.
Therefore the primordial tilt of $n_{s}$ can compensate $Y_{He}$
up to some extenct. In Fig.~\ref{fig:msoby} we have shown the
scatter plot between the scalar spectral index $n_{s}$ and the physical
baryon density $\Omega_{b}h^{2}$, and color coded the data
points according to the helium fraction i.e. $Y_{He}$. The one dimensional
marginalized distribution of primordial helium fraction is shown in
Fig.~\ref{fig:heliumlikelihood}, it can be seen that the likelihood
peaks close to the standard model helium fraction i.e. $0.24$. Therefore,
we can put a very tight constraint on the helium fraction from the
observational data from Planck. 

\subsection{SCoPE with full 19 Planck parameters}

The results from a simpler standard $6$ parameter model is presented in section(\ref{sec:6param})
with the nuisance parameters fixed 
to their average value obtained by the Planck collaboration \cite{Ade2013a}.
For testing the code it is important to run SCoPE on a higher dimensional 
parameter space. In this section we have SCoPE'd the 
$19$ dimensional parameter space with $6$ standard $\Lambda$CDM
parameters and other Planck nuisance parameters. With all 19 parameters the acceptance 
probability of the sample points is as low as  $\sim <1\%$. However, as SCoPE 
can run the indivudial chains in parallal, the acceptance probability can be 
increased as much as we want by increasing the number of processors. We have used 
$10$ CPU for parallelizing each of the chains.  
%Our results shows that even with $21$ parameters the Gelman-Rubin statistics
%converge just within first $~150$ steps that implies the mixing of the chains 
%are proper and the convergence is fast. 
and ensure that we take more than $~2000$ data points 
from each of the chain to get a better distribution of the posterior. 
The results of our analysis are in a good agreement with 
the Planck collaboration results. We have quoted the 
average and the best-fit values in Table \ref{tab:avgbest}.

{\tiny
\begin{table}
\centering
\begin{tabular}{|c|c|c|c|}
\hline 
{\tiny Parameter} & {\tiny Definitation} & {\tiny Mean $\pm$ SD} & {\tiny Best fit}\tabularnewline
\hline 
\hline 
{\tiny $\Omega_{b}h^{2}$} & {\tiny Physical baryon density} & {\tiny 0.0220$\pm$0.0003} & {\tiny 0.0220}\tabularnewline
\hline 
{\tiny $\Omega_{m}h^{2}$} & {\tiny Physical matter density} & {\tiny 0.1421$\pm$0.0028} & {\tiny 0.1424}\tabularnewline
\hline 
{\tiny $h$} & {\tiny Hubble parameter} & {\tiny 0.6712$\pm$0.0118} & {\tiny 0.670}\tabularnewline
\hline 
{\tiny $\tau$} & {\tiny Reion optical depth}  & {\tiny 0.089$\pm$0.013} & {\tiny 0.088}\tabularnewline
\hline 
{\tiny $n_{s}$} & {\tiny Scalar spectral index} & {\tiny 0.96$\pm$0.0068} & {\tiny 0.961}\tabularnewline
\hline 
{\tiny $\log(10^{10}A_{s})$} & {\tiny Scalar spectral amplitude}  & {\tiny 3.092$\pm$0.025} & {\tiny 3.126}\tabularnewline
\hline 
{\tiny $A_{100}^{PS}$} & {\tiny Contribution of Poisson point-source power to $D_{3000}^{100\times100}$
for Planck (in $\mu K^{2}$)} & {\tiny 183.0$\pm$52.9} & {\tiny 159.6}\tabularnewline
\hline 
{\tiny $A_{143}^{PS}$} & {\tiny Same as $A_{100}^{PS}$ but at $143$GHz} & {\tiny 56$\pm$11} & {\tiny 63.0}\tabularnewline
\hline 
{\tiny $A_{217}^{PS}$} & {\tiny Same as $A_{100}^{PS}$ but at $217$GHz} & {\tiny 113$\pm$13} & {\tiny 108.0}\tabularnewline
\hline 
{\tiny $A_{143}^{CIB}$} & {\tiny Contribution of CIB power to $D_{3000}^{143\times143}$ at the Planck
(in $\mu K^{2}$)} & {\tiny 10.26$\pm$3.353} & {\tiny 0.01}\tabularnewline
\hline 
{\tiny $A_{217}^{CIB}$} & {\tiny Same as for $A_{143}^{CIB}$ but for $217$GHz} & {\tiny 30.2$\pm$7.96} & {\tiny 44.6}\tabularnewline
\hline 
{\tiny $A_{143}^{tSZ}$} & {\tiny Contribution of tSZ to $D_{3000}^{143\times143}$ at $143$GHz (in
$\mu K^{2}$)} & {\tiny 6.1$\pm$3.54} & {\tiny 6.73}\tabularnewline
\hline 
{\tiny $r_{143\times217}^{PS}$} & {\tiny Point-source correlation coecient for Planck between $143$ and $217$GHz} & {\tiny 0.87$\pm$0.071} & {\tiny 0.902}\tabularnewline
\hline 
{\tiny $r_{143\times217}^{CIB}$} & {\tiny CIB correlation coecient for Planck between $143$ and $217$GHz} & {\tiny 0.42$\pm$0.23} & {\tiny 0.414}\tabularnewline
\hline 
{\tiny $\gamma^{CIB}$} & {\tiny Spectral index of the CIB angular power spectrum ($D_{l}\varpropto l^{\gamma^{CIB}}$)} & {\tiny 0.57$\pm$0.13} & {\tiny 0.625}\tabularnewline
\hline 
{\tiny $c_{100}$} & {\tiny Relative power spectrum calibration for Planck between $100$GHz and $143$GHz} & {\tiny 1.0005$\pm$0.00038} & {\tiny 1.0003}\tabularnewline
\hline 
{\tiny $c_{217}$} & {\tiny Relative power spectrum calibration for Planck between $217$GHz and $143$GHz} & {\tiny 0.9979$\pm$0.0013} & {\tiny 0.9983}\tabularnewline
\hline 
{\tiny $A^{kSZ}$} & {\tiny Contribution of kSZ to $D_{3000}$ (in $\mu K2$)} & {\tiny 4.98$\pm$2.62} & {\tiny 7.295}\tabularnewline
\hline 
\end{tabular}
\label{tab:avgbest}
\caption{Average, standard error and the best fit values of the parameters from the 
$19$ dimensional parameter estimation.}
\end{table}
}

\section{Conclusion and discussion}

We develop a new MCMC code named as SCoPE that can sample the posterior probability 
distribution more efficiently and economically than the conventional MCMC codes. In our code, the individual 
chains can run in parallel and a rejected sample can be used to locally modify the proposal 
distribution without violating the Markovian property. The latter increases the acceptance 
probability of the samples in chains. The prefetching algorithm 
allows us to increase the acceptance probability as much as required, provided 
requisite number of multiple cores are available 
in the computer. Apart from these, 
due to the introduction inter-chain covariance update the code can start 
without specifying any input covariance matrix. The mixing of the chains is also 
faster in SCoPE. 

The workability of the code is proved by analyzing different cosmological models. 
A 19 dimensional parameter estimation using SCoPE
shows that the method can be used to estimation the high dimensional 
cosmological parameters extremely efficiently.

\acknowledgments{S.D. acknowledge the Council of Scientific and Industrial Research (CSIR), 
India for financial support through Senior Research fellowships. T.S. acknowledges Swarnajayanti 
fellowship grant of DST India. We also like to thank Prof. Sanjit Mitra and Prakash Sarkar for 
their kind help during the project. 
Computations were carried out at the HPC facilities in IUCAA. }

\end{document}